# COMMENT ON
# MIE SCATTERING FROM A SONOLUMINESCING BUBBLE WITH HIGH SPATIAL AND TEMPORAL RESOLUTION

Physical Review E 61, 5253 (2000)

K.R. Weninger, P.G. Evans*, S.J. Putterman

Physics Department, University of California, Los Angeles CA 90095

*Division of Engineering and Applied Science Harvard University, Cambridge MA 02138

ABSTRACT: A key parameter underlying the existence of sonoluminescence [or 'SL'] is the time relative to SL at which acoustic energy is radiated from the collapsed bubble. Light scattering is one route to this quantity. We disagree with the statement of Gompf and Pecha that - highly compressed water causes the minimum in scattered light to occur 700ps before SL- and that this effect leads to an overestimate of the bubble wall velocity. We discuss potential artifacts in their experimental arrangement and correct their description of previous experiments on Mie scattering.

In sonoluminescence the first stage of energy focusing is provided by the collapse of a gas bubble surrounded by water. For example a 40KHz sound field with an amplitude of 1.35atm will cause a helium bubble with an ambient radius of 4microns to expand to a maximum radius of $29\mu$. Since the bubble contains $6.7 \times 10^9$ helium atoms and is acted on by about 1atm [=$P_0$] of ambient external pressure the mechanical energy stored when the bubble is perched at its maximum radius is about 10.eV per helium atom[1].

Sonoluminescence [SL] is due to some fraction of this potential energy being transferred into the thermal degrees of freedom of the relatively few atoms/molecules in the bubble as it implodes under the influence of $P_0$. For instance, if all the mechanical energy went into uniformly heating the helium, its temperature at the moment of collapse [when its radius approaches the van der Waals hard core of about $.4\mu$] would be about 75,000K.

A determination of the fraction of potential energy that remains in the bubble at the moment of collapse is a key aspect of SL. Thus the dynamics of the collapse is critical to an understanding of SL and so various techniques [2-10] have been applied to the experimental determination of the bubble radius as a function of time R(t) and the response of the water. Various realizations of Mie scattering [3-10] in particular have proved useful in obtaining bubble parameters. Mie scattering occurs when variations in the index of refraction cause light to be scattered out of the direction of the incident beam.

In a recent paper [11] Gompf and Pecha [GP] have used a streak camera to image Mie scattering. In the abstract they claim that "In the last nanoseconds around minimum bubble radius most of the light is scattered at the highly compressed water surrounding the bubble and not at the bubble wall. This leads to a minimum in the scattered light intensity about 700ps before the SL pulse is emitted." They go on to say that "neglect this changes [sic] leads to a strong overestimation of the bubble wall velocity".

We disagree with a number of aspects of these statements. The 700ps interval which GP quote is specific to their particular experimental arrangement and is unrelated to the physics of a bubble collapsing in highly compressed water. The stated timing resolution of GP is 500ps whereas by the application of time correlated single photon counting to Mie scattering we have achieved a timing resolution of about 50ps. In Figure 1 of reference [8] the flash will be seen to occur 100-200ps *before* [not 700ps *after*] the minimum apparent radius [i.e. 'y' axis], which for our experiment is the minimum in total light scattering. We agree [8] that light scattering is due to the index of refraction changes at the wall of the bubble as well as the highly compressed water. But in this case, attribution of the Mie scattering exclusively to the bubble wall would lead to an *underestimate*, of its velocity, not an *overestimate*, as quoted above from GP. Perhaps our observation that the flash *precedes* the minimum in light scattering could be due to this effect.

On page 5254 GP discuss our previous experiments on Mie scattering [3,4,5] and state that "In ..former investigations .. the scattered light intensity was assumed to be proportional to the square of the bubble radius which totally neglects the complicated angular distribution of the Mie scattering." This statement comprises an inaccurate description of past experiments. The complicated angular distribution can be seen in Figure 6 of [4], which was taken from our first paper on Mie scattering from SL [3]. One of the steps that enabled us to obtain quantitative information about bubble radii from Mie scattering was to simplify the scattered intensity as a function of R by collecting light from a large solid angle: such as $30°-80°$ [8] or $46°-94°$[3]. In this case the intensity of light scattering is within 20% of $R^2$ for bubbles bigger than, .6 microns[8], or 1micron[3]. These corrections and their connection to the "complicated angular distribution of Mie scattering" were discussed in these papers. A plot typical of calculations that formed the basis for these corrections was published in ref. [8]. The strong deviations from $R^2$ Mie scattering displayed in Figure 5 of GP results from their collecting light in a small solid angle $[14°-36°]$ near the forward direction. On page 5255 GP state that our papers neglected the effect of changes in the refractive indices. This is true; the index of refraction inside the bubble was reckoned to unity for the purpose of deconvolving the scattered intensity. Light scattering techniques have not yet reached the point where changes in the index of refraction, due to say the formation of a plasma, can be extracted. Analysis of our data also neglected the effect of bubble asphericity [see discussion relating to Figure 6 of ref.5].

The time scale of 700ps enters GP in two entirely different contexts: 1) it is a "pronounced minimum in the scattered light intensity .7ns before the SL pulse due to Mie lobe clusters" and 2) "From this time on most of the light is scattered at the highly compressed water around the bubble leading to a strong increase in the scattered light intensity before minimum bubble radius" which is the moment of SL. For the choice of angles over which GP collect scattered light we agree with 1) but emphasize that our choice of angles eliminated this artifactual minimum. Regarding 2) we reiterate our disagreement with GP's abstract.

At more than one location in GP it is claimed that "the bubble wall velocity 1ns before the SL pulse is about 950m/s. This value is much lower than the values found by Weninger et al "[reference 5 this comment]. First of all, the bubble wall velocity 1ns before collapse, where R is about $1.7\mu$, is for our data 900m/sec [5]. So it would appear that GP have confirmed our measurements. Secondly, the 500ps timing resolution of GP means that a 500ps smoothing function has been applied to their rapidly changing data. There can be no question that their value of 950m/s is an underestimate of the bubble wall velocity in their experiment. Furthermore, a statement claiming a significant discrepancy between experiments would have more weight if it was accompanied by a discussion of 'error bars'. The paper of GP contains no such discussion. We have attempted to give an example of our 'error' in Figure 6 of [5]. It has sources in the various processes discussed above, gas concentration, and also run to run variations.

The data of GP for R(t) cut off at about 1.7microns up to which point they are largely in agreement with our previously published results[4,5,6]. There remains the issue of whether the bubble wall velocity for some systems [4,5,6] approaches higher velocities [e.g. in excess of 1200m/s] for smaller radii. Systems are characterized by the gas mixture used, acoustic frequency and ambient temperature. [For 1%Xe 99%O dissolved at 150Torr driven at 40KHz at 20C, the maximum velocity was actually found to be less than 950m/s.(8)] In the range $1.7\mu > R > R_c \approx .5\mu$, where $R_c$ is the collapse or minimum radius, GP provide no data for R. They claim that this is due to the difficulty in subtracting out a large signal due to scattering from highly compressed water. Except for 200ps around the minimum we disagree. Based upon our experience we propose some possible complications that affect their experiment in this range; 1) the GP choice of angles leads to Mie lobes that are sufficiently complicated that the intensity of scattered light is not monotonic with radius, so that deconvolution is difficult, 2) as the GP images are magnified and averaged, small translational motion and concentric pulsation of the bubble can throw its image off the slit, 3) the level of scattered laser light is less than the intensity of SL obscuring dynamics near $R_c$. In Figure 1 we show photos of SL from a bubble undergoing a spontaneous motion on the order of $10\mu$. To the eye this particular bubble appears as stable as a star in the sky. When magnified by a factor of 10 this motion is enough to throw the image off the entry slit of the streak camera and introduce errors into the data. For integrated Mie scattering [3,4,5,6] where the width of the incoming light beam [typically 1mm] is larger than this motion and all the scattered light is delivered to a photo-detector the effects of this bubble motion can be minimized.  For this reason our investigations of imaged light scattering has been carried out on a shot by shot basis. In Figure 2 one sees a very common example of the bubble falling off the slit during that portion of the cycle that surrounds the minimum by about 1 ns as in GP. In Figure 2 the shrinking bubble is not centered on slit.

It is good news that the action of an audible sound field on water has led to a debate about experimental techniques on the scale of 100ps-700ps. As the time scale is narrowed it will eventually be possible to determine the time relative to SL '$t_a$-$t_{sl}$' at which acoustic energy is radiated by the bubble. This is very important because over 90% of the bubble's potential energy is radiated as sound[1]. If '$t_a$-$t_{sl}$' is negative then the average energy per particle at the moment of collapse is less than 1eV which is not high enough to explain the UV spectra [4] for a uniformly heated bubble[12]. Indeed if '$t_a$-$t_{sl}$' <0 the observation of SL would imply an additional energy focusing mechanism inside the collapsed bubble [13]. Perhaps the nonlinear processes that determine '$t_a$-$t_{sl}$' are the key to the existence of SL.

This research is supported by DARPA.

FIGURE CAPTIONS

1) A microscope image of light emission from a sonoluminescing bubble moving on the surface of a toroid of $15\mu$ radius. The bubble is formed from water which has a mixture of 1%He, 99%$O_2$ dissolved under a pressure of 150Torr. The acoustic frequency is 16.5KHz and the exposure time is .8seconds. In contrast B) shows our most stable image of SL (150Torr 1%Xe in $O_2$ at 41KHz) accumulated over 700flashes. The $1\mu$ resolution of the diameter is limited by both the microscope objective and possible bubble motion. A range of motions between those displayed in 'A' and 'B' may be seen under various circumstances which in general are beyond our experimental control. The intensity of SL is color coded so that red is brightest.

2) Single shot streak camera shadowgraph of a collapsing bubble launching a pulse of sound into the surrounding water [see ref.8 for experimental details]. The image of the bubble is the center line, and the radiated pulse of sound moves at a supersonic velocity relative to the speed of sound in water. The image of this particular bubble is lost during the indicated 1ns time-span. This effect is due to the bubble not being centered on the entrance slit of the streak camera. As the bubble shrinks its shadow falls off the slit. A photo lacking this artifact can be found in [8]. This experiment was carried out at 41KHz with 150Torr 1%Xe, 99%$O_2$.

A 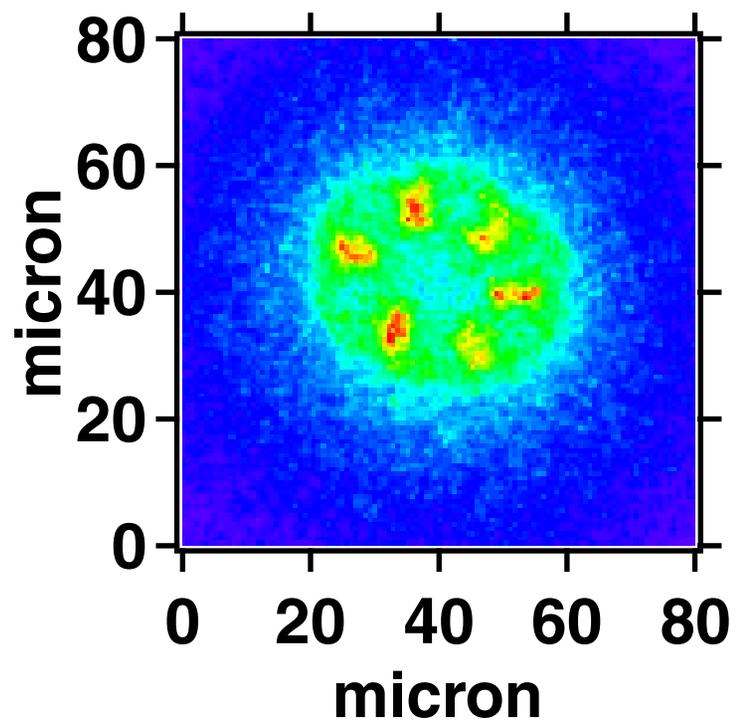

B 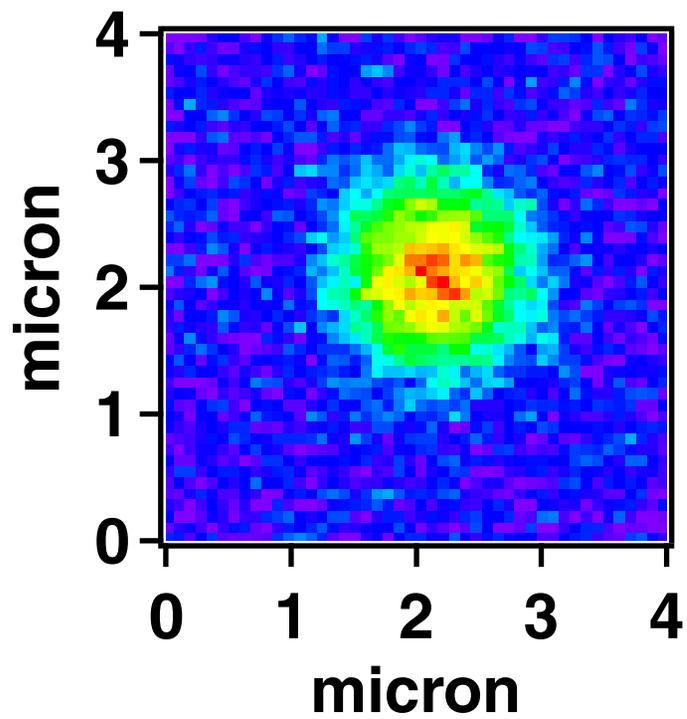

Figure 1

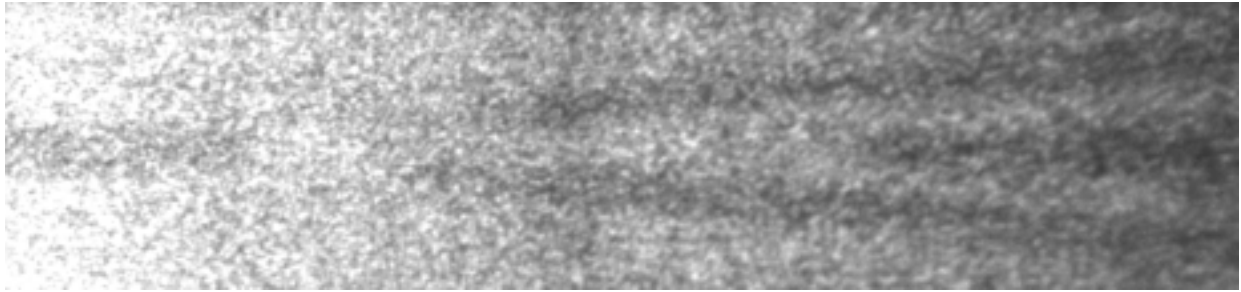

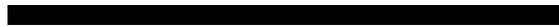

1 ns

Figure 2